# Decision-Making in a Social Multi-Armed Bandit Task: Behavior, Electrophysiology and Pupillometry

Julia Anna Adrian[1], Siddharth Siddharth[2], Syed Zain Ali Baquar[1], Tzyy-Ping Jung[3], & Gedeon Deák[1]

Department of Cognitive Science[1], Electrical Engineering[2], Bioengineering[3], UC San Diego
9500 Gilman Drive, La Jolla, CA 92093

**Abstract**

Understanding, predicting, and learning from other people's actions are fundamental human social-cognitive skills. Little is known about how and when we consider other's actions and outcomes when making our own decisions. We developed a novel task to study social influence in decision-making: the social multi-armed bandit task. This task assesses how people learn policies for optimal choices based on their own outcomes and another player's (observed) outcomes. The majority of participants integrated information gained through observation of their partner similarly as information gained through their own actions. This lead to a suboptimal decision-making strategy. Interestingly, event-related potentials time-locked to stimulus onset qualitatively similar but the amplitudes are attenuated in the solo compared to the dyadic version. This might indicate that arousal and attention after receiving a reward are sustained when a second agent is present but not when playing alone.

## Introduction

For successful social interaction it is useful to represent and predict other people's actions and the consequences of those actions. Joint action is defined as the ability to coordinate one's actions with others to achieve a goal (Vesper et al., 2016). Although it occurs in many sorts of human activities, it can be conveniently studied using the social or two-player versions of standardized cognitive tasks. Such tasks modified for social interactions can reveal complex dynamic in people's use of social information for judgments and action-planning.

For example, the *joint Simon task* (Sebanz, Knoblich, & Prinz, 2003) is a modified, two-player version of the standard Simon task that measures stimulus-response compatibility. Participants learn to respond with left or right button press for visual or auditory cues and show a longer reaction times are shorter when the cue location is compatible with the response hand, than when the cue occurs contralaterally. This *Simon effect* (Simon, 1990), interestingly, remains in the two-player version in which each participant is only responsible for one stimulus-response pair (Sebanz et al., 2003). This can be interpreted as evidence that human action planning is automatic and is elicited by processing another person's actions as well as planning and executing our own actions. The propensity to develop this ability might have evolved to enable efficient social learning (Kilner, Friston, & Frith, 2007; Liao, Acar, Makeig, & Deák, 2015). Particularly under conditions of uncertainty, the capacity to observe, encode, and imitate others' actions can be beneficial (Laland, 2004), permitting a sort of vicarious embodied modeling.

However, it is not always adaptive to generalize from other's actions and outcomes to one's own. The findings from joint Simon and other tasks have shown that representation of other's internal states occurs even when it is unnecessary, or disadvantageous, for optimal task completion. To study the extent to which people use observation of other's actions and outcomes to influence their own choices, even when it is unfavorable, we developed a novel task: the *social multi-armed bandit* task. The standard multi-armed bandit is a single-player paradigm to study decision-making under uncertainty. Named after the 'one-armed bandit' slot machines of casinos that have a fixed reward probability, multi-armed bandit tasks present several different options ('arms') of different, unknown reward probabilities. They manifest a classic exploitation/exploration problem (Cohen, McClure, & Yu; Gittins, Glazebrook, & Weber, 2011). Commonly, after an initial phase of exploration players employ one of two strategies: maximizing or matching. *Maximizing*, or consistently choosing the most-rewarding arm (based on prior observations), is the optimal strategy for problems with static reward probabilities. By contrast, *matching,* or choosing each arm in proportion to its relative reward probability, is suboptimal but nevertheless seen in humans and other animals (Sugrue, Corrado, & Newsome, 2004).

Notably, although a great deal of problem solving and prediction updating occurs in social or joint tasks, only a few studies have included multiple decision-makers in social versions of prediction tasks such as multi-armed bandit, and even these have not investigated effects of social interaction or observation on decisions (Liu & Zhao, 2010).

In addition to studying behavior, we recorded participants' electroencephalogram (EEG) and pupil size as physiological metrics of cortical and neuromodulatory concomitants of social decision-making. These bio-sensing methods may provide insights into the underlying neural dynamics of decision-making with high temporal resolution. Both of these physiological measures are common in affective computing (Partala, 2003) to measure valence and arousal, and cortical changes (Fink, 2009).

## The present study

The present study aims to address a *"key question of today's cognitive science: how and to what extent do individuals mentally represent their own and others' actions, and how do these representations influence, shape, and constrain an individual's own behavior when interacting with others?"* (Dolk et al., 2014).

To do so, we converted a classic three-armed bandit paradigm into a turn taking game. Reward probabilities for the three arms were different for each of two participants, allowing us to estimate the distinct effects of their own and their partner's action and outcome history on their ongoing decision-making. We studied three outcome measures: (1) decision-making behavior, (2) event-related EEG potentials, and (3) pupil dilation. Details are described below.

In the multi-player version, the probabilities remain constant for each player, however, they differ between the two players (see Table 1). This allows us to examine to what extent each participant takes into account their own and their partner's choices and outcomes. We expected to observe two different core strategies:

*Egocentric strategy:* Participants might make their decisions only based on their own outcome history and ignore information from their partner's outcomes. Players using this strategy should converge on choosing their own highest gaining arm (90% reward probability) most of the time.

*Joint strategy:* Participants might take into account information from their partner's outcomes to the same extent as information from their own outcomes. Players using this strategy should not converge on choosing one arm, because all arms average the same reward probability if both participants' outcomes are encoded equally. Alternately, in an intermediate strategy, participants might take into account their partner's outcomes but weigh them less than their own outcomes, and then more slowly converge on their own optimal choice.

Table 1: Reward probabilities for the different arms of the social multi-armed bandit

|          | Arm 1 | Arm 2 | Arm 3 |
|----------|-------|-------|-------|
| Player 1 | 30%   | 60%   | 90%   |
| Player 2 | 90%   | 60%   | 30%   |

# Experiment 1

## Participants

Participants were 28 female undergraduate students (14 dyads) recruited through the university's SONA system. They received course credit for participation in addition to a small monetary reward based on performance in the social multi-armed bandit (0.05 USD per reward).

One pair was excluded from EEG and eye-tracking analysis due to recording failure; another pair was excluded because one player chose the same arm on every trial. This left behavioral data from 26 participants, EEG data from 12 participants, and pupillometry data from 12 participants.

## Experimental Design

The (social) multi-armed bandit was described to participants as 'the ice-fishing game' and presented on a touchscreen. They were shown three 'ice holes' (arms) distinguished by shape, at approximately equal distances from each other (see Figure 2). The arms were associated with discrete and constant reward probabilities (30%, 60%, and 90%) unknown to the players. Upon choosing and touching a hole, participants heard and saw differential reward feedback. Participants had 100 trials each (200 total) to catch as many fish as possible, choosing one ice hole per trial. Each participant played the game once on their own (solo version) and once as a turn-taking game (dyadic version). For each dyad of participants, EEG and pupillometry data were collected from one player, and behavioral data were recorded from both.

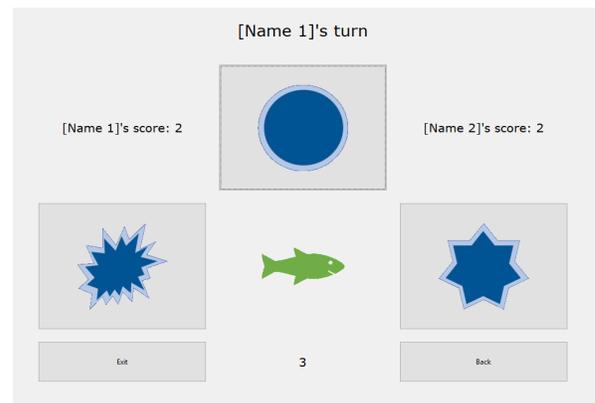

Figure 1: The game screen when a player has won a reward (green fish). The display shows the two players' accumulated rewards as well as which player's turn it is.

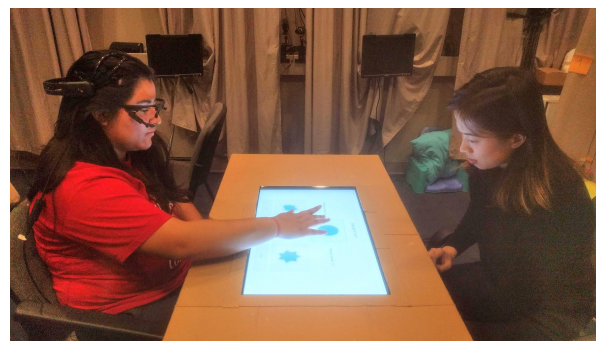

Figure 2: Two participants playing the social multi-armed bandit. The player on the left is wearing an Emotiv EEG headset and PupilLabs eye-tracker.

## Data Acquisition

The game was presented on a table-mounted capacitive touch screen monitor (diagonal: 66cm). During the dyadic turn-taking game the participants sat facing each other (figure 2). An Emotiv headset (www.emotiv.com) recorded 14-channel EEG data, and a PupilLabs headset (pupil-labs.com/pupil/) captured pupillometry data. These sensors were chosen for participants' comfort and natural movement during a social interaction. EEG was sampled at 128 Hz, and the eye-facing camera sampled at 120 Hz. PupilLabs Software was used to detect the pupil in each frame and calculate its diameter. Lab Streaming Layer (LSL) (Kothe, 2015) was used to synchronize all of the data streams (i.e. EEG, eye-gaze video, and game events) by time stamping each event and each sample.

Synchronized EEG and pupillometry data were locked to participants' game choices in LSL-created XDF files so that behavioral and physiological data were epoched to trials. On each turn the 2 sec of data following the outcome stimulus presentation (win/loss) was used for further analysis.

## Data Analysis

The first 20 trials of each game for each player were considered training trials, to teach the participant the game. These trials were not considered in the current analyses.

**Decision-Making Behavior** Participants' ice hole choice patterns were analyzed via Kullback-Leibler divergence (KLD), a measure of relative entropy.

$$D_{KL}(p, q) = \sum_{i=1}^{N} p(x_i) * log \frac{p(x_i)}{q(x_i)}$$

This quantifies the divergence of one probability distribution to another one. In our experiment we use the KLD to measure the difference between a participant's observed choices, and expected choices according to potential strategies. We hypothesize the employment of four different strategies with the expected choice probabilities as summarized in Table 2. For the egocentric-maximizing strategy, the player chooses the highest gaining arm (arm 3) at every trial. In the egocentric-matching strategy, each arm is chosen in proportion to their reward probability. The joint-equal strategy assumes that the outcomes of both players are weighted equally, resulting in an apparent reward probability of 60% for each arm. In that case each arm is chosen ⅓ of the trials. The joint-social strategy assumes a social value of the arm that has an equal, relatively high reward probability for both of the players (arm 2) and is thus chosen most often.

KLD of observed vs. expected probability distribution for each of the hypothesized strategies was calculated and compared to classify each participant's preferred strategy. As the joint strategy is not applicable in the single-player version, only the two egocentric strategies were compared.

Table 2: Expected choice probabilities for player 1 for each of the four hypothesized strategies. Reward probabilities for

| Strategy of player 1 | | Arm 1 | Arm 2 | Arm 3 |
|---|---|---|---|---|
| Ego-centric | maximizing | 0% | 0% | 100% |
| | matching | 16.7% | 33.3% | 50% |
| Joint | equal | 33.3% | 33.3% | 33.3% |
| | social | 0% | 100% | 0% |

**Game Data** The game data was an 8 x 200 matrix which included the turn number, player number, reward state, choice, time taken, player 1 reward and player 2 reward. In the single player case the last value was set as -1 and disregarded.

**EEG Data** The EEG data was cleaned using EEGLAB's Artifact Subspace Reconstruction (ASR) noise removal pipeline (Delorme, 2004; Mullen, 2013). Region of interest was the occipital cortex (channel O1).

**Eye-Tracking Data** The current analyses only consider a single channel containing pupil diameter information. Samples with abnormally high or zero pupil diameter values (due to detection errors or eye blinks) were ignored and data was interpolated by adjacent values.

After interpolating, the data was normalized to range between 0 and 1 to account for discrepancies in pupil diameter across subjects.

**Further Analysis** Epochs for each trial containing the response and 2 seconds of subsequent data (including the reward outcome). Our goal was to illustrate the pupil dilation (indicating autonomic response) and cortical dynamics (focusing on updating responses) upon perceiving a reward stimulus after choosing a specific action.

Pupil and EEG data for each type of choice and reward combination were then averaged across all subjects. For EEG data, each channel was averaged independently, to facilitate Event Related Potential (ERP) analyses. The 0.2 sec of data before the event were used as baseline for the normed succeeding EEG data, to control variance in EEG amplitude across subjects.

## Results

**Decision-Making Behavior** Over the course of the game, participants received information through trial and error and could learn that different arms were associated with different probabilities of receiving a reward. In the solo version, participants chose the highest gaining arm more often than the other two. Table 3 summarizes the decision-making behavior via mean total scores and mean number of choices for each arm. Participants distributed their choices more equally and scored lower during the dyadic game.

Table 3: Means (SD) of each decision type in the single- and multi-player games. All measures differ significantly between single and multi-player version (p<0.001).

|  |  | Reward probability | 30% | 60% | 90% |
|---|---|---|---|---|---|
| No. of choices by game version | single |  | 7 (5.0) | 17 (10.3) | 56 (13.4) |
|  | multi |  | 18 (9.1) | 32 (15.1) | 30 (19.5) |

Mean total score in the single-player game was 76 (SD: 7.0) compared to 63 (SD: 10.1) in the multi-player game (p<0.001).

For each version, participants were categorized based on the strategy employed. For each strategy, KLD was calculated between the observed and the expected choices. 70% (18/26) of participants employed a maximizing strategy in the solo version. In the dyadic version, strategies were more varied (see Figure 3). Most common was the joint-social strategy, used by 32% (9/28). There was no correlation between individuals' strategies in the solo and dyadic versions.

Strategy use affected overall scores in the dyadic version (F=6.083, p=0.004) and had a marginally significant effect in the solo version (p=0.073).

**Brain Dynamics** ERP locked to outcome stimuli for each of the differentially rewarding arms were compared between three conditions: (1) *solo*: a player's responses in the solo version of the game, (2) *dyad (self)*: a player's responses to an outcome of their own action, and (3) *dyad (other)*: a player's responses to an outcome of the partner's action.

Figure 4 illustrates the findings. The ERP displayed a prominent positive potential around 300ms (P3) after dyadic self-reward events in the dyadic version, but not for partner's reward or for reward in the solo game. We also note that the ERP response for partner's rewards as well as for own-reward in the solo game is attenuated but follows the same profile as that of the self-reward condition.

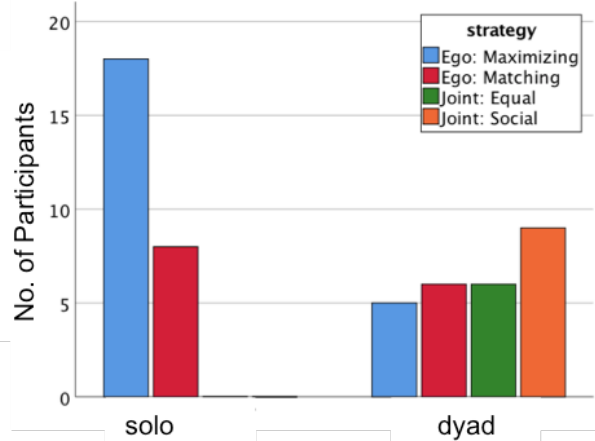

Figure 3: Distribution of strategies being employed by the participants in the single- and multi-player game, as determined via KLD.

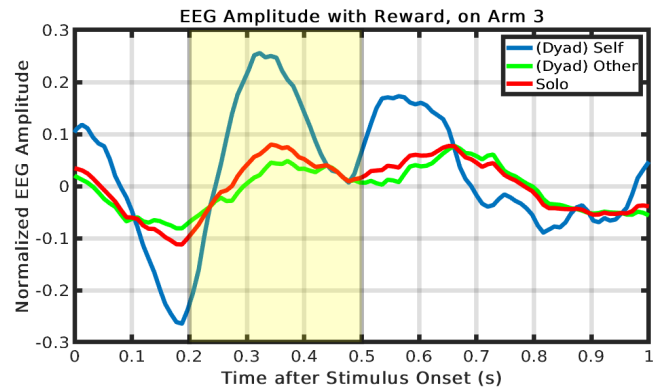

Figure 4: ERP after reward at arm 3 (90%/30%) averaged across participants from channel. Dyad (Self): player receives reward in dyadic version, Dyad (Other): player observes partner receive a reward in dyadic version, Solo: player receives a reward in solo version.

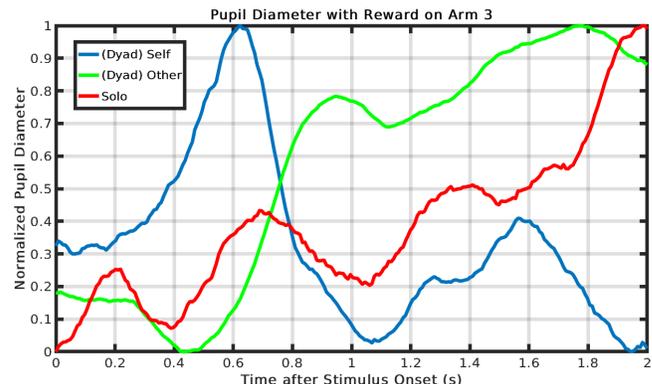

Figure 5: Pupillometry data after reward at arm 3 (90%/30%) averaged across participants from channel. Dyad (Self): player receives reward in dyadic version, Dyad (Other): player observes partner receive a reward in dyadic version, Solo: player receives reward in solo version.

**Pupillometry** Pupil dilation significantly increases initially (Figure 5) when a reward is obtained for the player themselves when playing the social multi-armed bandit, but not in the single-player version and not when observing the partner receive a reward. After this initial response, we see that after 0.8 seconds of the reward onset, pupil diameter increases for the Dyad (Other) condition whereas it decreases for the self-reward one.

## Discussion

The Social Multi-Armed Bandit revealed that adults take into account information from others' when making decisions. In consistency with previous studies, the majority of participants employed maximizing strategy in the solo version of the task. With this novel paradigm we found that this is not the case when there is a second player present, and when the reward probabilities for the arms are not the same for the two players. Instead, more than half of the participants employed a 'joint'-strategy, in which the actions and outcomes from the other player are integrated with their own when making decisions. One explanation for this phenomenon is a high prior belief of the same underlying probability structure for both players. This is likely the case because the visual representation remains constant throughout the game aside from updating the score count and the display whose turn it currently is. This issue is addressed in Experiment 2.

The ERP time-locked to stimulus onset showed a qualitatively similar pattern after receiving a reward at arm 3 (90%/30% reward probability) for the three conditions analyzed. However, the amplitude is highest when receiving a reward in the dyadic version, and attenuated when observing the partner receive a reward. When receiving a reward in the solo version of the game, the amplitude is also attenuated in comparison to receive a reward in the presence of a second player. We believe that this is due to the higher stakes and reward scenario attached with the dyadic version of the game.

Interestingly, pupil dilation increases drastically at about 0.6 seconds after stimulus onset when receiving a reward in the dyadic version of the game. In contrast, pupil dilation after observing the partner receive a reward has a longer latency of about 0.8 second. This likely reflects differential activation of the parasympathetic nervous system (PNS) for self/other reward scenarios.

# Experiment 2

## Participants

Participants were 32 undergraduate students (16 dyads, 10 female-female, 4female-male, 2 male-male) recruited through the university's SONA system. They received course credit for participation in addition to a small monetary reward based on performance in the social multi-armed bandit (0.05 USD per reward).

## Experimental Design

The experimental design of Experiment 2, is very similar to experiment 1. The modifications of the experiment are:

(1) Whereas in experiment 1, one person in each dyad played the solo version of the game before the dyadic version and the other person played in the reverse order, in experiment 2 both played the solo version either before or after the dyadic version. In other words, game order was randomly assigned by dyad. This ensured that the game process was not driven by prior knowledge of only one of the players.

(2) To reduce the prior belief of a constant underlying reward structure of the game, we changed the background color of the game after each turn, such that there was a distinct visual cue to signal each player's turns.

(3) EEG data was recorded from both participants in Experiment 2 (vs. only one participant per dyad in Exp. 1). Pupillometry data was not recorded.

## Data Acquisition

See Experiment 1.

## Data Analysis

As in Experiment 1, the first 20 trials were excluded from analysis.

### Decision-making behavior

The analysis performed in Experiment 1 is based upon the assumption that participants make use of particular strategies. In this experiment, a different type of analysis was performed, considering choice behavior 'bottom-up' without assumptions of specific strategies.

Participants' choices were analyzed via the Jensen-Shannon Divergence (JSD). The JSD is a distance metric between two probability distributions and based on the KLD:

$$D_{JS}(p,q) = \frac{1}{2}D_{KL}(p,x) + \frac{1}{2}D_{KL}(q,x)$$

with $x = (p+q)/2$

The JSD between the relative choice distribution of the last 80 trials of the participants' empirical behavior and the relative choice distribution if all choices were made towards the highest gaining arms (=(0,0,1), maximizing strategy) was used to analyze the data. Hence, the decision-making behavior for each participant could be characterized by their JS divergence in the solo and the dyadic version. As reference, the JS divergence of relative choice distribution between matching behavior (0.17, 0.33, 0.5) and maximizing (0,0,1) is 0.31. This value was used to further cluster participants into 'learners' (JSD < 0.31) and 'non-learners' (JSD ≥ 0.31).

## Results

**Decision-Making Behavior** Participants could be categorized into four groups, depending on if they learned which option was the highest gaining in the solo and/or dyadic version of the game. 50% (16/32) of participants were grouped into 'learners' in the solo version, but into 'non-learners' in the dyadic version. As shown in Figure 6, in the solo version, they choose the 90% reward arm significantly more often than the other two arms (F = 179.1, p < 0.0005) and significantly more often than in the dyadic version (F = 59.7, p < 0.0005). 22% (7/32) of participants were clustered into 'learners' in both versions of the game, and 19% (6/32) were clustered into 'non-learners' in both versions of the game. 9% (3/32) were clustered into 'learners' in the dyadic version of the game, but not in the solo version.

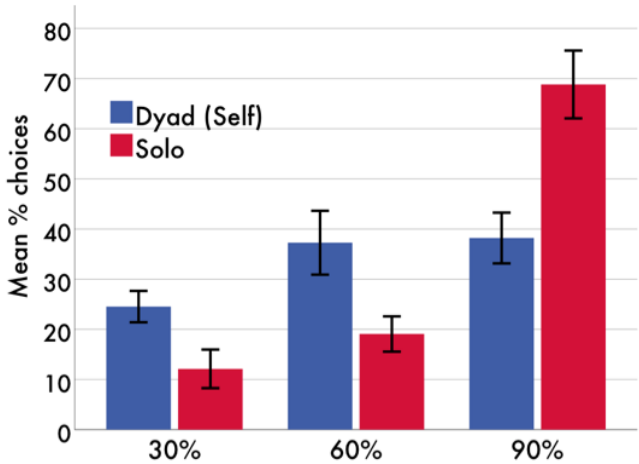

Figure 6: Choice behavior of 50% of participants who learned which arm has the highest reward probability in the solo but not the dyadic version of the multi-armed bandit.

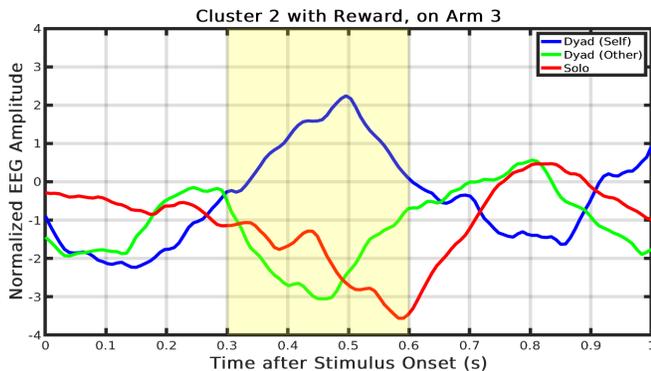

Figure 7: ERP after reward at arm 3 (90%/30%) averaged across participants from channel. Dyad (Self): player receives reward in dyadic version, Dyad (Other): player observes partner receive a reward in dyadic version, Solo: player receives a reward in solo version.

**Brain Dynamics** ERPs were examined at channel O1 of participants who chose the highest gaining arm most of the time in the solo version, but not in the dyadic version (Figure 7). We observed a high increase in amplitude for the Dyad (Self) condition. In comparison, there was a negative deflection in ERP for the Dyad (Other) and Solo condition.

## Discussion

Even when the prior belief of a common underlying reward structure was decreased, half of the participants integrated information gained through observation of their partner similarly as information gained through their own actions.

In Experiment 2, our goal was to combine the subjects' decision-making behavior with their physiology. Similarly as in Experiment 1, the ERP time locked to stimulus onset when receiving a reward at arm 3 showed a high increase in amplitude for the Dyad (Self) condition. In comparison, there was a negative deflection in ERP for the Dyad (Other) and Solo condition. We consider this a good starting point to move towards extracting more high-level features such as EEG power spectrum density, mutual information and pupil diameter-based fixations and saccades in the future.

## General Discussion

We developed a Social Multi-Armed Bandit task to examine the influence of social interaction on decision-making. We found that while some individuals do figure out that the other player's information does not apply to them, the majority of participants converged to a suboptimal decision-making strategy. We termed this strategy 'joint' as it most likely emerges through averaging the reward probabilities for both players. Measurement of electrophysiology showed a distinct P3 when the player receives a reward in the dyadic version of the multi-armed bandit but not the solo version. P3 is thought to emerge through stimulus-driven 'top-down' processes when the participant pays focused attention to a task. The distinct presence of the P3 in the dyadic task might thus hint towards heightened attention, particularly towards own rewards, in the presence of another player. Interestingly, the pupillometry data revealed a similar pattern as the ERP.

This task has considerable possibilities for further studies of social interaction. Next steps include a similar experiment with participants are previously acquainted with each other, e.g. friends, and children with their parents. It is likely that having a prior relationship with the other partner will alter the joint strategy for one or both partners. It would also be interesting to test how an asymmetric relationship (e.g., parent-child) would influence decision-making strategies, compared to a more symmetric relationship. It is possible that less-experienced participants (e.g., children) are more likely to follow, or match, the behavior of a 'reliable' person, as is the case in imitation (Poulin-Dubois, Brooker, & Polonia, 2011). Lastly, this task might give interesting insights into decision-making processes of neuro-divergent people, particularly those with potential differences in social behaviors (Montague, 2018).

## Acknowledgements

This study was funded with an Innovative Research Grant from the Kavli Institute for Brain and Mind.